\documentclass[pra,aps,twocolumn,10pt,superscriptaddress,footinbib]{revtex4-1}
\usepackage{amsmath,amssymb,bbm,bm}
\usepackage{graphicx}
\usepackage{epstopdf}
\usepackage{latexsym}
\usepackage[normalem]{ulem} 
\usepackage[usenames,dvipsnames]{color}
\usepackage{hyperref}
\usepackage{natbib}
\usepackage{physics}
\usepackage{bm}
\usepackage{xcolor, colortbl}
\begin{document}

\newcommand{\expt}[1]{\langle #1 \rangle}
\renewcommand{\mod}[1]{\lvert #1 \rvert}
\newcommand{\modsq}[1]{\mod{#1}^2}
\newcommand{\ns}{\mathcal{N}_{\mathrm{s}}}
\newcommand{\sinc}{\mathrm{sinc}}
\newcommand{\nb}{\mathcal{N}_{\mathrm{b}}}
\newcommand{\warn}[1]{{\color{red}\textbf{* #1 *}}}

\newcommand{\eqplan}[1]{{\color{blue}\textbf{equations:{ #1 }}}}

\newcommand{\figplan}[1]{{\color{blue}\textbf{figures: { #1 }}}}

\newcommand{\tableplan}[1]{{\color{blue}\textbf{tables: { #1 }}}}

\newcommand{\warntoedit}[1]{{\color{blue}\textbf{EDIT: #1 }}}

\newcommand{\warncite}[1]{{\color{green}\textbf{cite #1}}}

\newcommand{\Rev }[1]{{\color{black}{#1}\normalcolor}} 
\newcommand{\Com}[1]{{\color{red}{#1}\normalcolor}} 
\newcommand{\LBCom}[1]{{\color{blue}LB: {#1}\normalcolor}} 

\bibliographystyle{apsrev4-1}

\newcommand{\mytitle}{The Effect of Micromotion and Local Stress in Quantum simulation with Trapped Ions in Optical Tweezers}
\newcommand{\rg}[1]{{\textcolor{purple}{#1}}}

\newcommand{\affA}{QuSoft, Science Park 123, 1098 XG Amsterdam, the Netherlands}
\newcommand{\affB}{Institute for Theoretical Physics, Institute of Physics, University of Amsterdam, Science Park 904, 1098 XH Amsterdam, the Netherlands}
\newcommand{\affC}{Van der Waals-Zeeman Institute, Institute of Physics,
University of Amsterdam, 1098 XH Amsterdam, Netherlands}

\title{\mytitle}
\date{\today}

\author{Liam Bond}\affiliation{\affA}\affiliation{\affB}
\author{Lisa Lenstra}\affiliation{\affC}
\author{Rene Gerritsma}\affiliation{\affA}\affiliation{\affC}
\author{Arghavan Safavi-Naini}\affiliation{\affA}\affiliation{\affB}

\begin{abstract} 
The ability to program and control interactions provides the key to implementing large-scale quantum simulation and computation in trapped ion systems. Adding optical tweezers, which can tune the phonon spectrum and thus modify the phonon-mediated spin-spin interaction, was recently proposed as a way of programming quantum simulators for a broader range of spin models [Arias Espinoza et al., Phys. Rev. A {\bf 103}, 052437]. In this work we study the robustness of our findings in the presence of experimental imperfections: micromotion, local stress, and intensity noise. We show that the effects of micromotion can be easily circumvented when designing and optimizing tweezer patterns to generate a target interaction. Furthermore, while local stress, whereby the tweezers apply small forces on individual ions, may appear to enable further tuning of the spin-spin interactions, any additional flexibility is negligible. We conclude that optical tweezers are a useful method for controlling interactions in trapped ion quantum simulators in the presence of micromotion and imperfections in the tweezer alignment, but require intensity stabilization on the sub-percent level.
\end{abstract}

\maketitle  

\section{Introduction}

Trapped ions are at the forefront of both digital and analog quantum simulation~\cite{Cirac:1995, Porras_2004, Blatt:2008}. On the digital side, trapped-ions are the building blocks of the highest fidelity two-qubit universal gates~\cite{Brown:2011, Ballance:2016, Gaebler:2016}, and the recent demonstration of on-the-fly quantum error correction adds to the robustness of this architecture~\cite{Bohnet:2021}. On the analog side, they have been used to emulate the dynamics and prepare the ground states of quantum magnets, as well as study the dynamics of quantum correlations, quantum information and entanglement in the presence of engineered, variable-range interactions~\cite{Kim:2009, Monroe_Correlations_2014, Monroe_MBL_2016, Rey_MQC_2017, Bermudez:2011, Bermudez:2011}.

Trapped-ion quantum simulators allow one to engineer power-law spin-spin interactions which decay as $1/r^\alpha$ where $0<\alpha<3$ and $r$ is the distance between two ions. This is the direct result of the mechanism behind the interactions. The inter-ion interactions are phonon-mediated and as such depend on the spectrum and structure of the collective vibrational modes of the ion crystal~\cite{Britton:2012, Freericks:2015}. So far experimental efforts utilizing trapped-ions as analog simulators have been restricted to the aforementioned power-law interactions. 
Recently it has been shown that the addition of optical tweezers to the typical trapped-ion platform produces a highly tunable quantum simulator in terms of connectivity, range, and sign of the interactions in both linear (or 1D) and triangular (2D) ion crystals in Paul traps~\cite{Espinoza:2021, Teoh:2021, Nath:2015, Olsacher:2020}. If a target interaction matrix passes our feasibility criterion, we search for the optimal optical tweezer pattern to manipulate the frequencies and structure of the collective vibrational modes of the crystal. 

In this work we study the robustness of our scheme in presence of typical experimental imperfections: micromotion, tweezer misalignment, and tweezer intensity noise. In Section.~\ref{sec:trapped-ion-qsim} we review the radio-frequency (r.f.) Paul trap and the formalism describing the motion (including micromotion) of ion crystals. In Section.~\ref{sec:micromotion-spinspin-engineering} we extend previous studies to characterize the effect of small-amplitude micromotion~\cite{Landa:2012, Kaufmann2012, Duan:2015} and correct for it in our tweezer patterns, before including first-order Doppler modulation. Section.~\ref{sec:LocalStress} investigates if local stress due to misalignment of the tweezers can improve the optimization and considers the effect of laser intensity fluctuations. 

\section{Trapped-ion quantum simulator}\label{sec:trapped-ion-qsim}
We consider a one or two dimensional crystal of $N$ ions in a Paul trap. The potential energy of the system is given by $V_0 = V_{\rm coulomb}+V_{\rm trap}$. The first term is the contribution due to the Coloumb repulsion between the ions $V_{\rm coulomb}(\bm{r}_i)=\frac{1}{2} \sum_{i\neq j} \abs{\bm{r}_i - \bm{r}_j}^{-1}$, whilst the second term is the confinement supplied by the external trapping potential
\begin{align}
    V_\text{trap}(r_{i,\alpha},t) = \frac{\Omega_\text{rf}^2}{8} \sum_{i,\alpha} [a_\alpha - 2 q_\alpha \cos(\Omega_\text{rf} t)] r_{i,\alpha}^2,  \label{eq:generalTrappotential}
\end{align}
generated by DC fields and AC components oscillating at $\Omega_{\rm rf}$. Here  $a$ and $q$ are the (dimensionless) Mathieu parameters and $r_{i,\alpha}$ is the position of the $i$-th ion in the  $\alpha=x,y,z$ direction. The ion positions and the oscillation frequency are dimensionless and in terms of the characteristic length scale $d = \left(e^2/(4\pi\epsilon_0 m \bar{\omega}^2)\right)^{1/3}$ and a characteristic frequency $\bar \omega$ respectively. Here $e$ is the electron charge, $\epsilon_0$ is the vacuum permittivity and $m$ is the ion mass. This allows us to define time $t$ in units of $1/\bar{\omega}$. Thus Eq.~\ref{eq:generalTrappotential} is dimensionless with an energy scale $m \bar{\omega}^2 d^2$. 

The interplay between the external trapping potential and the Coulomb repulsion results in stable Coulomb crystals. The dimensionality of the crystal depends on the relative strength of the trapping potential along the different axes \cite{Dubin_1993,Enzer_2000}. We focus on the case of a 2D zigzag crystal in the $yz$-plane, as shown in Fig.~\ref{fig:pseudopotentialVSmicromotion}(a). Tight confinement along $x$ ensures the crystal forms in the $yz$-plane, whilst a weaker potential along $z$ compared to $y$ (or vice-versa) leads to the formation of the zigzag structure.

The equilibrium positions of the ions are given by the solutions to $\nabla V_0 = 0$. The full solution is equilibrium positions with explicit time-dependence $\mathbf R_{i}(t)$ to account for micromotion even at ultra-low temperatures. However when $\abs{a},q^2 \ll 1$ we make the pseudopotential approximation and replace the time-dependent potential $V_\text{trap}$ with a static harmonic potential~\cite{James:1998}
\begin{align}
    V_\text{pseudo}(r_{i,\alpha}) = \frac{1}{2}\sum_{i,\alpha} \Theta_\alpha^2 r_{i,\alpha}^2, \label{eq:PPTrapPotential}
\end{align}
where $\Theta_\alpha = \gamma_\alpha \Omega_\text{rf}/2$ are effective frequencies determined by the characteristic exponents of the Mathieu equation, $\gamma_\alpha \approx \sqrt{a_\alpha + q_\alpha^2/2}$ \cite{McLachlan1947}. Note that although the Mathieu exponents are usually denoted by $\beta$, we use $\gamma$ to avoid confusion with a later use of $\beta$. 

The emergence of effective spin-spin interactions, mediated by the collective oscillations (phonon modes) of the crystal have been previously studied. The phonon-mediated interactions are generated by applying a spin-dependent force, using a Raman beam pair, to couple the electronic spin of the ion to the collective motion of the crystal. Within this approximation trapped-ion quantum simulators allow us to engineer spin-spin interactions that decay as $1/r^{\xi}$, with $0\leq \xi \leq 3$ \cite{Richerme2016, Britton:2012, Kim:2009, Porras_2004}. The interaction strength between ions $i$ and $j$ is given by 
\begin{align}
    J_{i,j} = \sum_m \frac{(\bm{k} \cdot  \bm{b}_{i,m})(\bm{k} \cdot\bm{b}_{j,m})}{\mu^2 - \omega_m^2}, \label{eq:spinspincoupling}
\end{align}
where $\bm{b}_{i,m}$ is a $3$-element vector (each element describing a direction $\alpha$) of the $m$-th mode and the $i$th ion, $\bm{k}$ the $3$-element wave vector of the Raman beam pair, $\omega_m$ the frequency of the $m$-th mode and $\mu$ the Raman beat-note frequency. Thus the structure of the spin-spin interactions is fully determined by the normal modes of the crystal and the beat-note frequency $\mu$. Here we have assumed that the phase of the Raman beam pair driving the side-band transitions remains constant at the equilibrium position of the ions. 

In the absence of any additional control knob, one is limited to the power-law interactions described above. We have previously shown that a wider variety of target spin-spin interactions can be engineered by modifying the mode structure with optical tweezers \cite{Espinoza:2021}. We assume that the tweezers have cylindrical symmetry and supply confinement in the yz-plane only. We also assume that the micromotion amplitude is sufficiently small such that each ion stays near the center of the tweezer beam, and that the tweezer beam is focused on the ion equilibrium positions $\bm{R}_{i}$. Then the tweezer potential can be written as a local harmonic potential for each ion, 
\begin{align}
    V_\text{tweezer}(r_{i,\alpha}) = \frac{1}{2} \sum_{i=1}^{N}\sum_{\alpha = y,z}\nu_i^2 (\tilde{r}_{i,\alpha})^2,
\end{align}
where $\nu_i$ is the pinning frequency on the $i$th ion and $\bm{\tilde{r}}_i = \bm{r}_{i} - \bm{R}_{i}$ are the ion positions relative to their equilibrium. In the pseudopotential approximation the equilibrium positions are natively time-independent; when including micromotion we average the time-dependent equilibrium positions over one r.f. period. We denote the total potential, including tweezers, by $V_\text{total} = V_\text{trap} + V_\text{coulomb} + V_\text{tweezer}$.

\subsection{Equilibrium Positions with Micromotion} 
When optical tweezers are added to the system, in principle the solution to $\nabla V_\text{total} = 0$ gives the equilibrium positions. However for simplicity we assume that the equilibrium positions are unaffected by the tweezer potentials, which we justify in Section.~\ref{sec:micromotion-spinspin-engineering} by showing that our engineered coupling matrix is unaffected by this approximation. 

The equilibrium positions are thus given by the solution to $\nabla V_0 = 0$. We set the characteristic frequency $\bar{\omega} = \Omega/2$, and re-scale time accordingly $t \rightarrow \Omega t / 2$ to make the micromotion $\pi$-periodic. The $3N$ coupled equations of motion (eoms) are then \cite{Leibfried:2003}
\begin{align}
    \ddot{r}_{i,\alpha} + [a_\alpha - 2 q_\alpha \cos(2 t)] r_{i,\alpha} - \sum_{i\neq j} \frac{ r_{i,\alpha} - r_{j,\alpha}}{\abs{\bm{r}_i - \bm{r}_j}^{3}}  = 0.
    \label{eq:ionEOMs}
\end{align}
The addition of a cooling term $V_{\rm cool} = f(t) \dot{\bm{r}}_{i}$, where $f(t)$ is a time-dependent cooling profile that ramps from $f(0)=1$ to $f(t_{\rm max})=0$ allows us to start from an initial guess and evolve to the equilibrium configuration at $t_{\rm max}$. We then evolve the positions for one more period with $f(t>t_{\rm max})=0$ to determine the time-dependent equilibrium positions $\bm{R}_{i}(t)$. 


\subsection{Linearized Motion} 
To calculate the normal mode structure we follow the steps in Refs.~\cite{Landa:2012,Kaufmann2012}. We linearize the eoms about small oscillations of the equilibrium positions $\bm{\tilde{r}}_{i} = \bm{r}_{i}-\bm{R}_{i}$, 
\begin{align}
    \ddot{\tilde{r}}_{i,\alpha} + [a_\alpha - 2q_\alpha \cos(2t)] \tilde{r}_{i,\alpha} + \sum_{j,\beta} D_{i,j}^{\alpha,\beta}(t) \tilde{r}_{j,\beta} = 0,
\end{align}
where the time-dependent Hessian is defined as
\begin{align}
    D_{i,j}^{\alpha,\beta}(t) = \left. \frac{\partial^2 V_{\rm Coulomb}}{\partial r_{i,\alpha} \partial r_{j,\beta} } \right \vert_{r_{i,\alpha} = R_{i,\alpha}}. \label{eq:Hessian}
\end{align}
The linearized eoms have periodic coefficients and thus can be treated using Floquet theory. Expanding the Hessian matrix in a Fourier series as
\begin{align}
D = D_0 - 2 D_2 \cos(2t) - \hdots
\end{align}
where the matrices $A$ and $Q$ are defined as $A = {\rm diag}(a) + D_0$ and $Q = {\rm diag}(q) + D_2$, the matrix $\Pi(t)$ and vector $\bm{\phi}$ are introduced as
\begin{align}
\Pi(t) = \begin{pmatrix}
0 & \mathbbm{1} \\ 
-A + 2Q\cos(2t) & 0
\end{pmatrix}, \quad
    \bm{\phi} = \begin{pmatrix}
    \tilde{r}_{i,\alpha} \\ \dot{\tilde{r}}_{i,\alpha}
    \end{pmatrix},
\end{align}
where $\mathbbm{1}$ is a $3N$-dimensional identity matrix. The linearized eoms are then written as linearly independent equations in $6N$-dimensional phase space as
\begin{align}
    \dot{\bm{\phi}} = \Pi(t) \, \bm{\phi}. \label{eq:floquet}
\end{align}
We solve the set of differential equations to obtain the Floquet modes and exponents, which are related to the eigenmodes $\bm{b}_m^{f}$ and eigenfrequencies $\omega_m^{f}$ of the linearized ion-crystal motion (using superscript $f$ to denote that the solutions are from the full motion treatment). 

To obtain the eigenmodes and eigenfrequencies in the pseudopotential approximation we construct the Hessian as defined in Eq.~\ref{eq:Hessian}, but where the partial derivatives are now with respect to the static equilibrium positions $\bm{R}_{i}$. The Hessian is therefore time-independent and can be simply diagonalized to yield the eigenmodes $\bm{b}_m^{p}$ and eigenfrequencies $\omega_m^{p}$ (using superscript $p$ to denote the pseudopotential solutions). 

\subsection{Micromotion of a 2D Zigzag Crystal}\label{subsec:EffectMicromotion2DResults}
To characterize the effect of micromotion we study a $N=12$ ion crystal using experimentally relevant trap parameters. Specifically we use $a = \{0.018704, -0.018900, 0.000196\}$, $q = \{0.202780 , -0.202780 , 0 \}$ and $\Omega_\text{rf} = 2\pi\times 20 \text{ MHz}$. The corresponding pseudopotential frequencies are $\Theta_\alpha = 2\pi \times \{2,0.4,0.14\}\text{ MHz}$.

\begin{figure}
    \centering
    \includegraphics[width=1\linewidth]{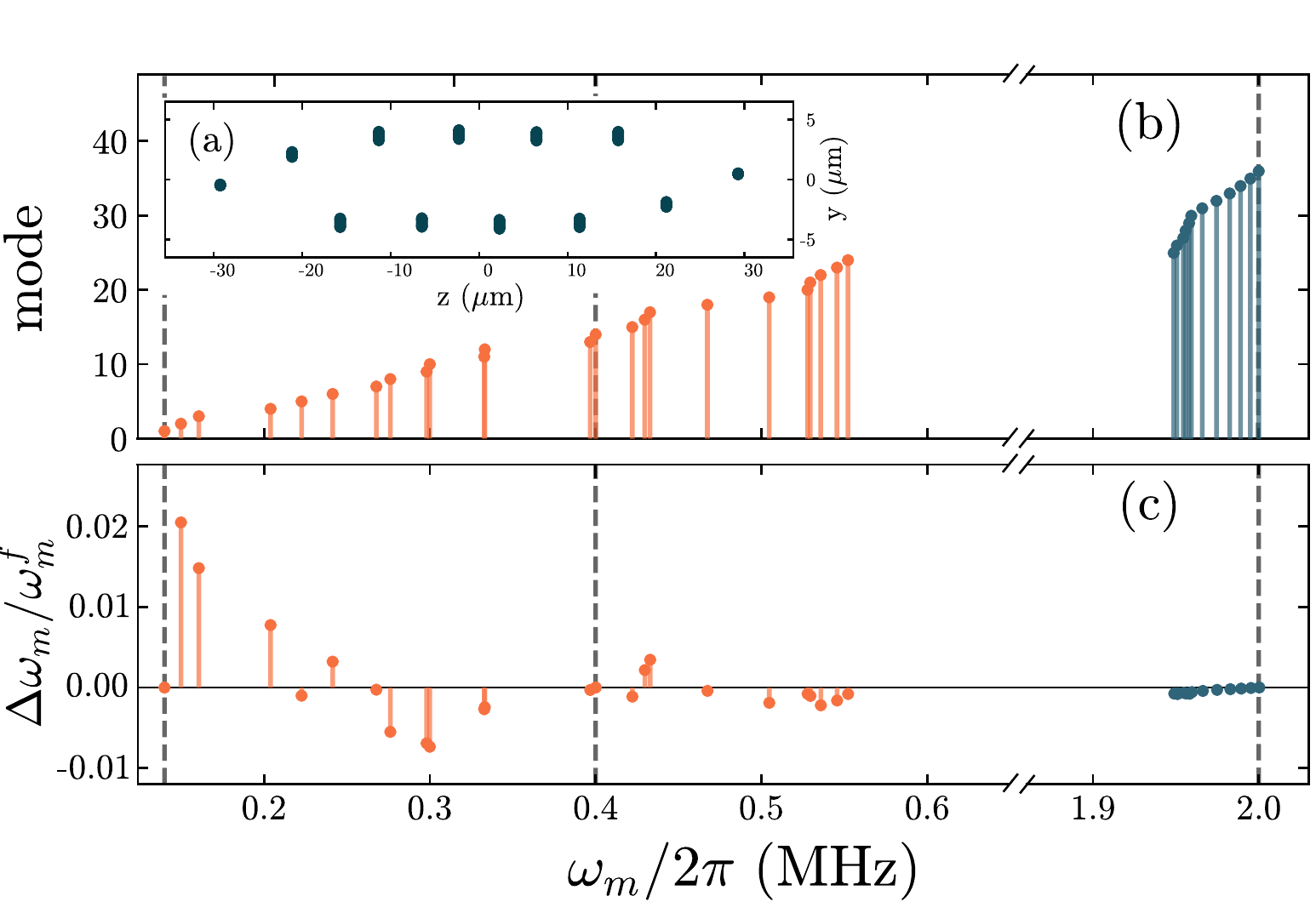}
    \caption{The effect of micromotion on a $N=12$ ion zig-zag crystal. (a) Ion positions during one r.f. period with motion indicated by blurring. Micromotion occurs only in the $y$ direction. (b) Mode frequency spectrum for $yz$ plane (orange) and $x$ (blue). The com mode frequencies (vertical black dashed lines) are unchanged by micromotion. (c) Frequency shift $\Delta \omega_m$ normalized to $\omega_m^f$. All shifts are small (kHz) relative to the mode frequencies themselves (MHz).}
    \label{fig:pseudopotentialVSmicromotion}
\end{figure}

Fig.~\ref{fig:pseudopotentialVSmicromotion}(a) shows the ion equilibrium positions with blurring to indicate micromotion over one r.f. period. Micromotion occurs only in $y$ with amplitude proportional to the ion's distance from the $y = 0$ trap axis, as described by the first order approximation $(1/2)q_{\alpha} R_{i,\alpha}$. In Fig.~\ref{fig:pseudopotentialVSmicromotion}(b) we plot the spectrum. Because the micromotion is a breathing mode oscillation the center of mass (com) modes are unchanged. The out-of-plane modes (along $x$) are decoupled from the in-plane modes ($y$ and $z$) and have a higher frequency and smaller bandwidth. Fig.~\ref{fig:pseudopotentialVSmicromotion}(c) shows the frequency shift $\Delta \omega_m = \omega_m^{f} - \omega_m^{p}$ normalized to $\omega_m^f$. Although the frequency shift is larger for modes with more breathing or zigzag-like structure, the frequency shifts are all relatively small $(\text{kHz})$ compared to the mode frequencies themselves $(\text{MHz})$. As such, from the mode structure itself we conclude that the pseudopotential approximation is justified. 

\section{Engineering spin-spin interactions in Optical Tweezers}\label{sec:micromotion-spinspin-engineering}
In this section we investigate if micromotion restricts our ability to engineer a target spin-spin interaction. We demonstrate that although tweezer patterns determined in the pseudopotential approximation are unsuitable once micromotion is included, corrected tweezer patterns can be found. However, the Doppler shift of the laser implementing the spin-spin interactions does cause an appreciable degradation in the engineered interaction compared to the target which is challenging to correct.



\subsection{Naive Inclusion of Micromotion}
We firstly make the pseudopotential approximation and numerically optimize the tweezer frequencies $\nu_i$ and Raman beat-note frequency $\mu$ to engineer a target coupling matrix. To characterize the success of the optimization, we define an error function as
\begin{align}
    \epsilon = \frac{\norm{J_T - J_E}}{\norm{J_T}}, \label{eq:errorfunc}
\end{align}
 where $J_E$ and $J_T$ are the engineered and target interaction matrices respectively, and where the matrix norm is the Frobenius norm.

During the optimization we assume that the equilibrium positions are unchanged by the tweezers. To justify this approximation we find that applying a maximum $2\pi \times 10 \text{ MHz}$ tweezer frequency on all ions causes an ion position change of $\sim 10 $~nm, and that using the corrected ion positions with an optimal set of tweezer frequencies causes a negligible change in $\epsilon$ on the order of $10^{-3}$. 

\begin{figure}
    \includegraphics[width=\linewidth]{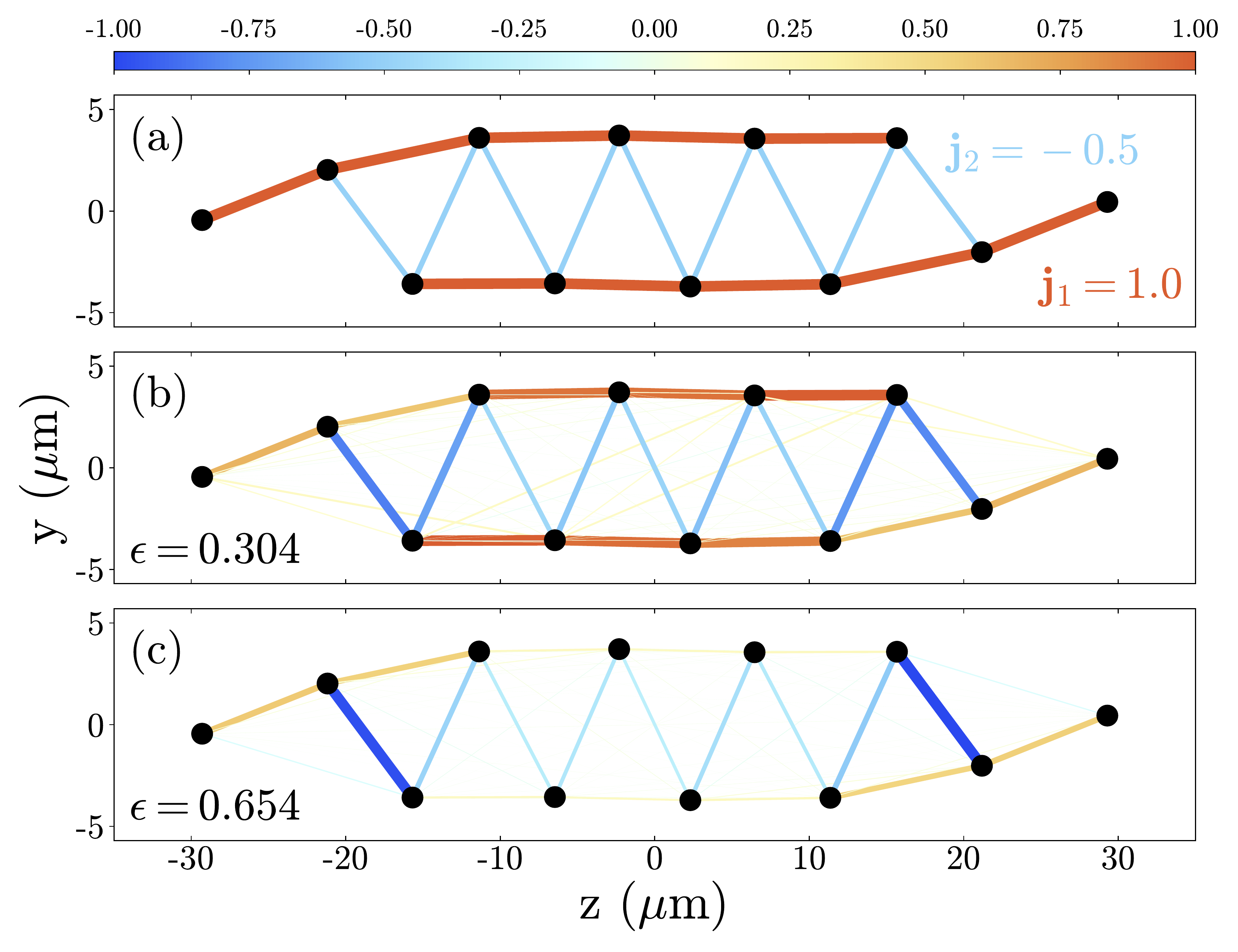}
\caption{(a) Spin-spin couplings for the target zig-zag coupling $J_T$ with $- j_2/j_1 = 0.5$. (b) Engineered couplings $J_E$ in the pseudopotential approximation. With optical tweezers, the target coupling can be engineered with reasonably low error. (c) ``Naive'' inclusion of micromotion by using the tweezer parameters found in the pseudopotential case. The difference in mode structure results in a large increase in $\epsilon$, making the tweezer solution found in the pseudopotential approximation unsuitable.}
\label{fig:mmppresults}
\end{figure}

For the target coupling we use a spin-ladder interaction, as shown in Fig.~\ref{fig:mmppresults}(a). Here we choose the spin-ladder since it is challenging to realize in ion crystals utilizing only the collective modes of the crystal in the absence of the tweezer potentials. It also offers variety via the coupling strength ratio $j_2/j_1$ enabling us to study the interplay of frustration and fluctuations, necessary ingredients for spontaneous continuous or discrete symmetry breaking in condensed matter systems. The ability to tune the range of zig-zag coupling strengths ($|j_2/j_1| \gg 1$) will allow us to study the phase diagram of this well-known frustrated magnetic system with no exact solution. 

To perform the numerical optimization we use Simulated Annealing, implemented using \emph{Optim.jl} \cite{mogensen2018optim} version 1.6.1 in \emph{Julia} \cite{bezanson2017julia} version 1.6.2.  We limit the maximum tweezer laser power to 30~W and use beam waists of $w = 1 \,\mu\text{m}$. The tweezer frequencies are upper-bounded by $\nu_i/(2\pi) \leq 1.0 \text{ MHz}$ whilst the Raman transition frequency is bounded by $0.3 \text{ MHz} \leq \mu/(2\pi) \leq 1.0 \text{ MHz}$. In addition we demand that $|\mu - \omega_m| > 10 \text{ kHz} $ to ensure the phonon modes  are only excited virtually. We implement this final requirement in the optimisation routine by adding a large value to the cost function defined in Eq.~\ref{eq:errorfunc} if the condition is not satisfied. 

Fig.~\ref{fig:mmppresults}(b) shows the optimal interaction graph and corresponding error $\epsilon_p = 0.304$ that can be realized in the pseudopotential approximation. In Fig.~\ref{fig:mmppresults}(c) we ``naively'' take the optimal tweezer pattern found in the pseudopotential approximation and recalculate the error using the micromotion equilibrium positions and mode structure, finding  $\epsilon_m = 0.654$. The difference $\epsilon_{m}-\epsilon_{p} = 0.350$ is significant, with the interaction graph showing little spin-ladder structure. As such, any optimization should include micromotion during the routine. 

\subsection{Including Micromotion during optimization}
Including micromotion during the optimization routine requires re-calculating the time-dependent Hessian with a given set of $\nu_i$ and solving the $6N$ Floquet equations to find the new mode structure. Although this procedure is computationally costly, for larger $N$ the cost can be reduced by using the symmetry of the coupling matrix in the tweezer patterns. For example, the spin-ladder interaction is symmetric about $z = 0$ and thus the tweezer frequencies can be assumed to obey the same symmetry. For the $N=12$ Coulomb crystal we find this is not necessary, and so optimize over all $12$ tweezer frequencies. 

\begin{figure*}
    \includegraphics[width=\linewidth]{{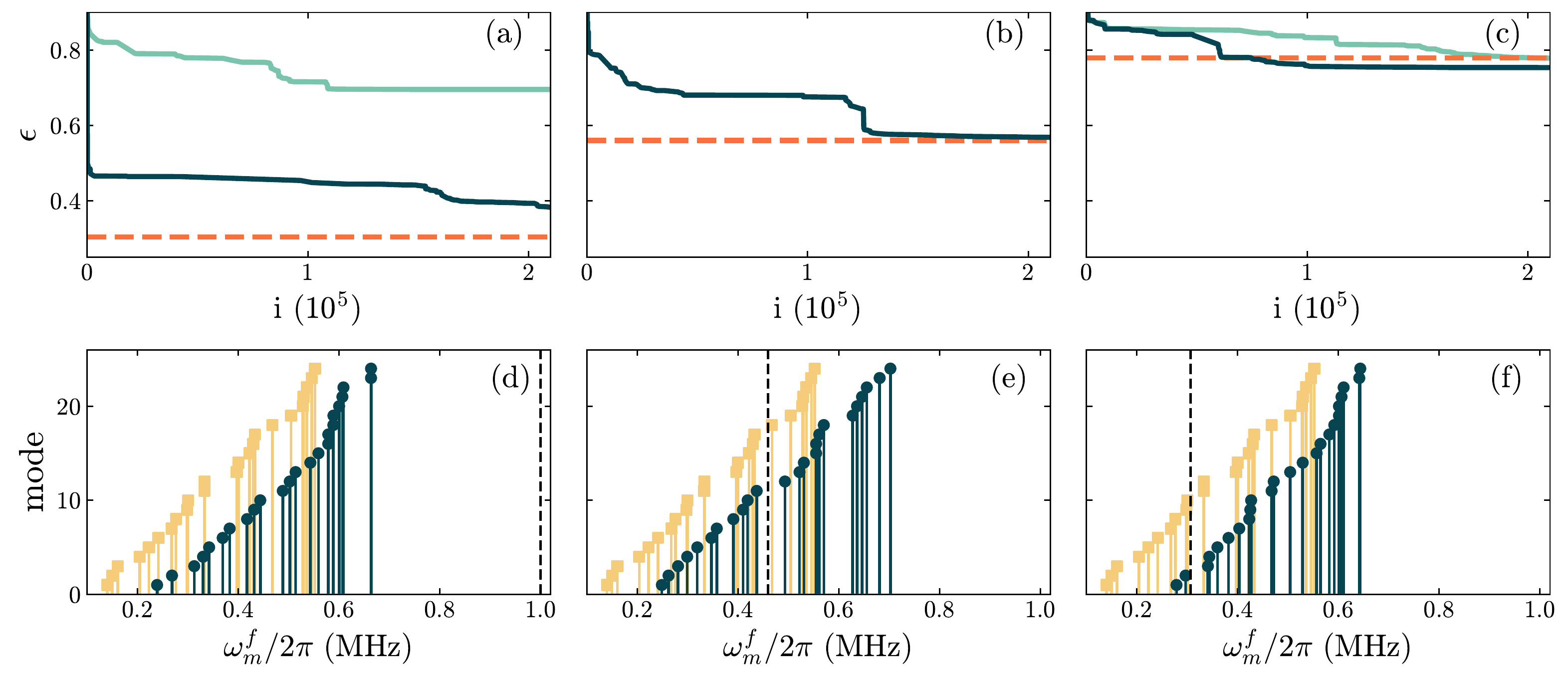}}
    \caption{Panels (a), (b) and (c) show the error $\epsilon$ as a function of optimization evaluations ($i$) for  wave vectors $\bm{k} = [0,1,0]$, $\bm{k} = [0,0,1]$ and $\bm{k} = [0,1,1]$ respectively. The target coupling is the spin-spin ladder shown in Fig.~\ref{fig:mmppresults}. When including micromotion in the optimization (dark blue line) we obtain a similar $\epsilon$ as the pseudopotential case (dotted orange line). The $\text{k} = [0,1,0]$ case (panel a) shows the best performance due to the tighter confinement along $y$.  Modulation has a detrimental effect in this scenario, as this is the direction where the micromotion amplitude is largest. Panels (d), (e) and (f) show the native spectrum (yellow) and tweezer-modified spectrum (dark blue) corresponding to (a), (b) and (c) respectively. The black dashed line shows the optimized beatnote frequency $\mu$. Note that $\abs{\omega_m^f - \mu} < 10 \text{kHz}$ to maintain a dispersive spin-phonon coupling. }
    \label{fig:optimresults}
\end{figure*}

In Fig.~\ref{fig:optimresults} we plot the optimization of $\epsilon$. When micromotion is included in the optimization, $\epsilon$ approaches the pseudopotential result. As such, micromotion itself is not a significant barrier to engineering interactions with optical tweezer.

\subsection{First Order Doppler Modulation}
The first order Doppler shift can have a significant impact on the spin-spin couplings. Following the procedure used in Ref.~\cite{Berkeland:1998} to lowest order in $a$ and $q$ the laser field (up to a phase factor) in the reference frame of the moving ion is
\begin{align}
E_i(t) = \Re \left[ \bm{E}_0 e^{i \bm{k}^ \cdot \bm{r}_{i}} \sum_{n = -\infty}^{\infty} \mathcal{J}_n(\beta) e^{-i \omega t + i n (\Omega_\text{rf} t)} \right],
\end{align}
where $\omega$ is the frequency and $\bm{k}$ the wave vector, and $\mathcal{J}_n(\beta_i)$ the Bessel function. The (dimensionless) modulation index $\beta_i$ is given by
\begin{align}
\beta_i = \frac{1}{2} \abs{\sum_{\alpha} k_\alpha R_{i,\alpha} q_\alpha}.
\end{align}
The carrier transition amplitude is modified by $\mathcal{J}_0(\beta_i)$, and thus the interaction matrix element becomes 
\begin{align}
J_{i,j}^\text{doppler} = \mathcal{J}_0(\beta_i) \mathcal{J}_0(\beta_j) J_{i,j}, 
\end{align}
where $J_{i,j}$ is the unmodulated coupling matrix element given in Eq.~\ref{eq:spinspincoupling}. Assuming a $411 \text{nm}$ laser we include Doppler modulation in the optimization. The resulting $\epsilon$ is shown in Fig.~\ref{fig:optimresults}. Although there is no Doppler modulation in $z$ (because $q_z = 0$) nor $x$ (because $R_{i,x} = 0$), there is significant modulation along $y$. The reduction in coupling strength depends on the distance of each ion from the $y = 0$ r.f. null, which makes it challenging to correct for using optical tweezers. While this ion-dependent source of error can be compensated by tuning the intensity of the Raman beams on each ion, the extra infrastructure cost is prohibitive.

\section{Local Stress}\label{sec:LocalStress}
In Section~\ref{sec:micromotion-spinspin-engineering} we used tweezer beams centered on the average equilibrium positions of the ions to more accurately engineer spin-ladder interactions. However if the tweezer beams are offset from the equilibrium positions, the tweezers add not only a local trapping potential but also supply a force. In this section we investigate if this local stress enables further improvements to our engineered couplings. We show that tweezer offsets of up to $0.25\,\mu\text{m}$ offer only small improvements to $\epsilon$. 

\subsection{First Order Approximation}
For simplicity we assume that we have a geometry in which micromotion does not play a role. As before we assume that the tweezers have cylindrical symmetry and supply confinement in the $yz$-plane only. The tweezer potential including an offset is then given by 
\begin{equation}
V_{\text{tweezer}}(r_{i,\alpha}) = \frac{1}{2}\sum_{i=1}^{N}\sum_{\alpha=y,z} \nu_i^2 (\tilde{r}_{i,\alpha}-\delta r_{i,\alpha})^2,
\label{v_tw}
\end{equation}
where $\bm{\tilde{r}}_{i} = \bm{r}_{i} - \bm{R}_{i}$ are the positions of the ions relative to their equilibrium, $\bm{\delta r}_{i}$ is the tweezer offset from $\bm{\tilde{r}}_i$ and the characteristic frequency is now set to $\bar{\omega}=\Theta_z$. 

Offsetting the tweezers changes the equilibrium positions of the ions. To find the new equilibrium positions $\mathbf{R}_i + \bm{\rho}_i$ we need to solve $\nabla_{\mathbf{\tilde{r}}} V_\text{total} =0$. This is computationally costly for large crystals, particularly when included in an optimization routine. Instead, as a first approximation we assume that the tweezers pull lightly on the ions, $\bm{\rho}_{i}/\bm{\delta r}_{i} \ll 1$. This is equivalent to treating the tweezers as a small perturbation compared to the Paul trap and Coulomb interactions. For simplicity we omit the $x$-direction, which is justified when the laser implementing the spin-spin interactions has no effective wave vector in the $x$-direction and the sound wave modes in the $x$-direction decouple, such as in a 2D ion crystal in the $yz$-plane. These prerequisites can be easily obtained by design. Denoting the Hessian matrix of $V_0 = V_\text{trap} + V_\text{coulomb}$ by $D_0$, we expand $\nonumber \nabla_{\mathbf{\tilde{r}}} V_{\text{tot}}(\mathbf{\rho})$ to first order,
\begin{align}
    \nonumber \nabla_{\mathbf{\tilde{r}}} V_{\text{tot}}(\mathbf{\rho}) &\approx  \left(\nabla_{\mathbf{\tilde{r}}}\left(\nabla_{\mathbf{\tilde{r}}} V_{0}(\mathbf{\tilde{r}})\right)\right)_{\mathbf{\tilde{r}}=0}\text{ }\bm{\rho}+\bm{\nu}^2(\bm{\rho} - \bm{\delta r})\\    
    &= D_0(0)\bm{\rho}+\bm{\nu}^2(\bm{\rho} - \bm{\delta r}),
    \label{notwsimp}
\end{align}
where $\bm{\nu}$ is a $2N\times 2N$ diagonal matrix with diagonal elements $\nu_i$. Note the zeroth order term drops out since $(\nabla V_0)_{\mathbf{\tilde{r}}=0}=0$ by definition. The lowest order shifts in the equilibrium positions are therefore
\begin{equation}
    \bm{\rho} \approx \left(D_\text{tot}(0)\right)^{-1}\bm{\nu}^2 \delta \bm{r},
    \label{neweqpos}
\end{equation}
where $D_\text{tot}(0)=D_0(0) + D_{\text{tw}}$ and $D_{\text{tw}}=\bm{\nu}^2$. 

Having approximated the new equilibrium positions, we now calculate the change in the Hessian matrix. To avoid calculating the Hessian $D_{\text{tot}}(\bm{\rho})$ directly from the new potential, we use an approximation to further reduce the computational cost,
\begin{equation}
    D_{\text{tot}}(\bm{\rho})\approx D_\text{tot}(0) + \left(\nabla_{\mathbf{\tilde{r}}}(D_0)\right)_{\mathbf{\tilde{r}}=0}\text{ }\bm{\rho}+\hdots. 
    \label{dapprox}
\end{equation}
$D_{\text{tot}}(\bm{\rho})$ has new eigenfrequencies $\tilde{\omega}_m^{\text{str}}$ and eigenvectors $\tilde{\mathbf{b}}_m^{\text{str}}$ resulting in new spin-spin interactions as defined by Eq.~(\ref{eq:spinspincoupling}). Although only approximate, this equation gives insight into the effect of the local stress on the mode spectrum. Because both $D_{\text{tw}}$ and $D_\text{trap}$ are constant diagonal matrices, the derivatives of $D_\text{tot}(0)$ originate from the Coulomb interaction alone. Due to the long-range character of the Coulomb interactions we expect that the local stress should ease the simulation of long-range interactions. On the other hand, the local stress terms are higher order than the tweezer curvature terms, so we expect the capability of local stress to significantly change the mode spectrum to be limited. Although this suggests local stress will not offer improvements to our engineered couplings, the benefit is that errors due to misaligned tweezers are suppressed. 

\subsection{Optimization}

\begin{figure}
    \centering
    \includegraphics[width=1\linewidth]{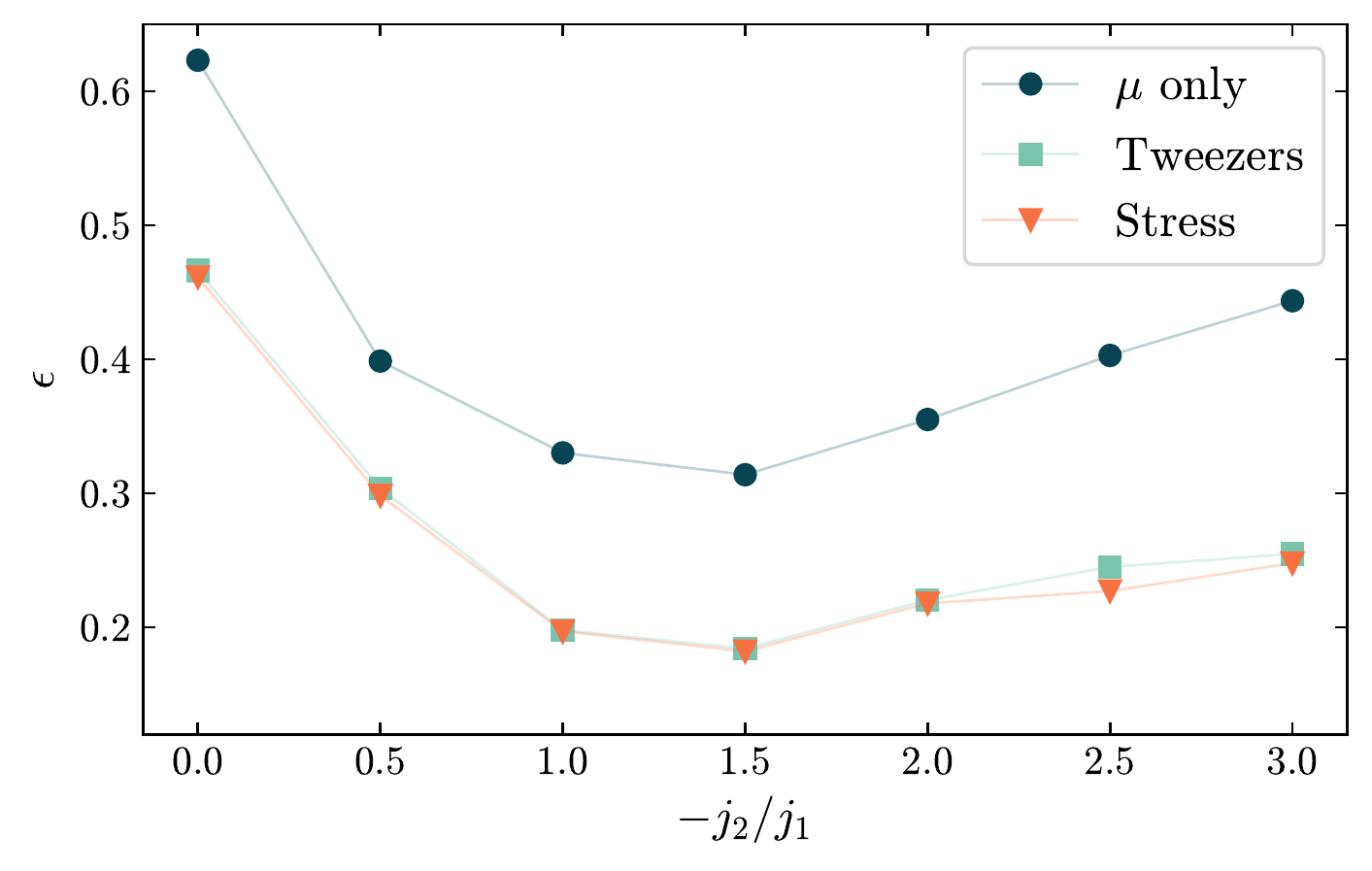}
    \caption{Error $\epsilon$ as a function of the spin-ladder coupling strength ratio $-j_2/j_1$. The error is smallest at $-j_2/j_1 = 1.5$ as this most closely resembles power-law interactions that can be well-engineered natively. The addition of tweezers offers significant improvement at all ratios. As predicted by our approximate expression Eq.~\ref{dapprox}, local stress of up to $0.25\,\mu\text{m}$ only offers a small improvement. This suggests that the couplings are robust to tweezer misalignments. }
    \label{fig:spinladderstressvsnostress}
\end{figure}

We investigate numerically whether it is possible to improve on the results obtained in the previous section if we allow the tweezers to supply local stress on the ions. For the $N=12$ ion crystal we fix the tweezer pattern to the optimal solution found in Sec.~\ref{sec:micromotion-spinspin-engineering} and optimize the tweezer offsets $0 \leq \bm{\delta r_i} \leq 0.25\,\mu\text{m}$. The offset bounds enable us to approximate the tweezers as harmonic. By fixing the tweezer parameters, we only need to optimize over the $2N$ offset parameters, and therefore in the optimization routine can calculate the new equilibrium positions $\bm{R}_i = \bm{\rho}_i$ and Hessian directly. Note that optimization over the full parameter set (including the tweezer parameters) is possible, particularly with a two-step optimization routine that firstly uses the approximate calculations of Eqs.~\ref{neweqpos} and \ref{dapprox} to determine if the parameters are promising, and then when the error falls below a set threshold uses the exact calculation to fine-tune the parameters and obtain the true error. We also optimize the full parameter set in this manner and find no difference to our fixed-tweezer optimization. 

In Fig.~\ref{fig:spinladderstressvsnostress} we vary the ratio $-j_2/j_1$ in the $12$-ion spin-ladder and calculate the error as defined in Eq.~\ref{eq:errorfunc}. As expected, the inclusion of tweezers results in significant improvements in engineering the target spin-spin interactions. However applying local stress to the ion crystal only results in minimal improvements. As such we conclude that in the perturbative regime local stress offers little benefit, but is reassuring since the interactions are therefore robust to tweezer misalignments. 

\subsection{Intensity noise}\label{sec:IntensityNoise}
\begin{figure}[t!]
    \includegraphics[width=1\linewidth]{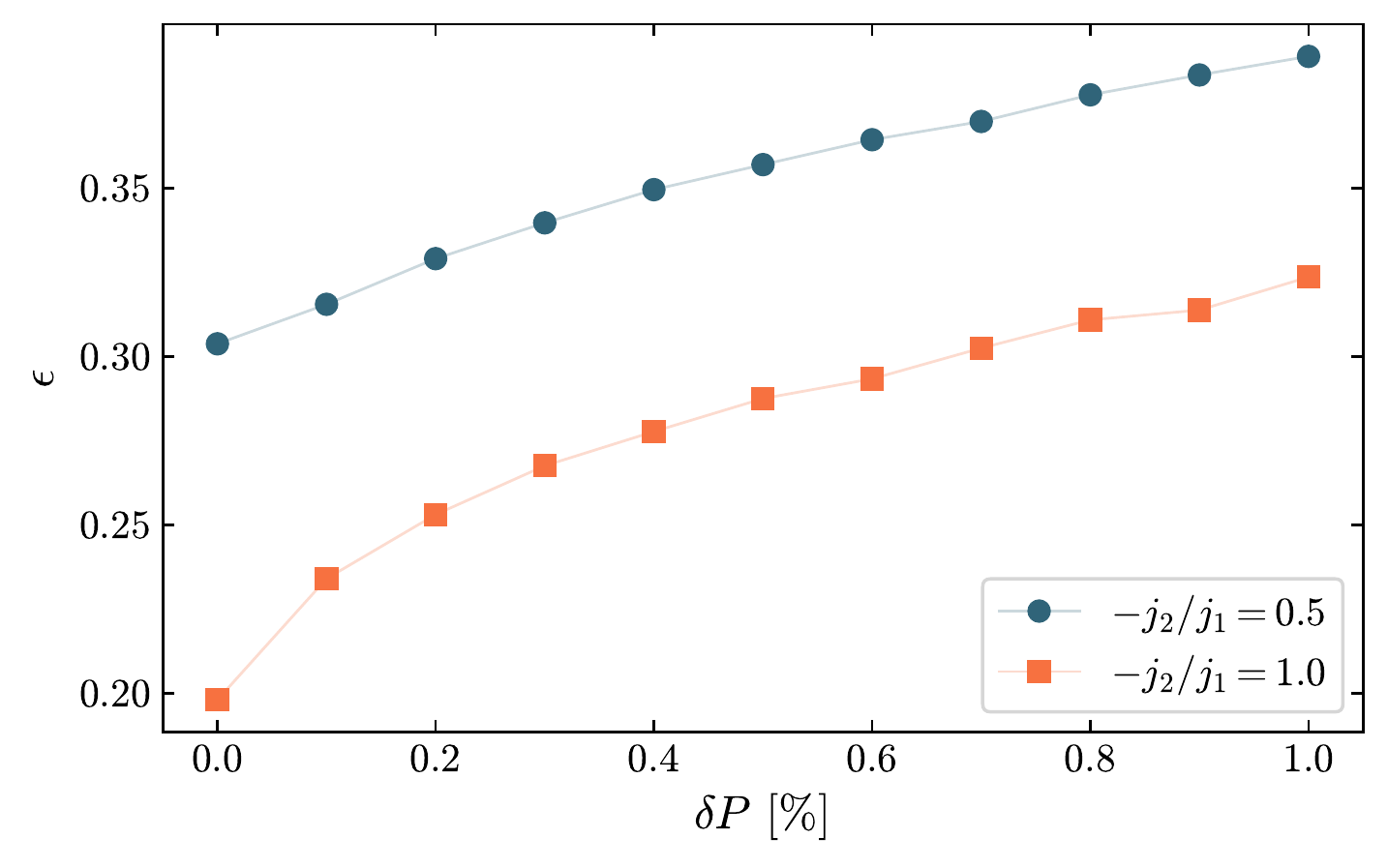}
    \caption{Error $\epsilon$ when tweezer intensity fluctuations $\delta P$ are included for two different spin-ladder coupling strength ratios. The error is calculated assuming random Gaussian noise in the laser generating the optical tweezers at frequencies much slower than the coupling time.}
    \label{fig:spinladdertweezerfluctuations}
\end{figure}

Finally, we study the effect of tweezer intensity fluctuations. We consider a worst-case shot-to-shot noise scenario, whereby an optimal set of tweezer frequencies $\nu$ are each subject to a fluctuation $\delta \nu$. Note that $\delta \nu \propto \sqrt{\delta P}$, where $\delta P$ is the power fluctuation, since the square of the tweezer frequencies are proportional to the laser power. To simulate the noise we multiply an optimal tweezer pattern by a random fluctuation sampled from a normal distribution with standard deviation $\delta P$. We repeat the calculation $N_\text{repeat} = 10^4$ times and take the average. In Fig.~\ref{fig:spinladdertweezerfluctuations} we plot $\epsilon$ as a function of the percentage noise in the tweezer power $\delta P$. We find that for typical experimental parameters intensity noise on the order of $\lesssim 1\%$ can have a noticeable impact on the engineered coupling. As such, intensity stabilization on the order of sub-percent is required to accurately engineer the target spin-ladder coupling.

\section{Conclusions}\label{sec:Conclusion}

Local optical potentials, supplied by optical tweezers, allow us to create analog trapped-ion quantum simulators with an unprecedented level of flexibility concerning the possible spin-spin interaction patterns. In this work we studied the robustness of this approach in a typical experimental setup. In particular, we focused on three sources of error: (i) micromotion, (ii) tweezer misalignment, and (iii) tweezer intensity noise.  We used the ferromagnetic zig-zag model, with $j_1>0$ and $j_2<0$, to quantify the adverse effect of each source of error. Our choice of model is motivated by the fact that tweezers play a fundamental role in generating the target connectivity and the range of interactions. Hence this model provides us with a upper bound on the sensitivity of the scheme to the three sources of error listed above.

We showed that the effect of micromotion is two-fold. First, it shifts the motional modes of the crystal, and second, it causes a first-order Doppler shift and in turn modulates the spin-spin couplings for each ion. We showed that the shift in the motional modes is at the level of few percent, justifying the use of the pseudopotential approximation. However the first-order Doppler shift may be a major source of error along the weaker confinement direction when micromotion is the largest. In contrast, we find that in the limit where the tweezer potential is perturbative compared to the Paul trap and the Coulomb interactions, any additional stress and strain force on the ions due to the misalignment of the tweezers is negligible. Finally we find that the intensity noise should be controlled to the sub-percent level, as this shot-to-shot noise severely impacts the fidelity with which the target interactions can be realized.

\begin{acknowledgments}
We thank Juan Diego Arias-Espinoza for sharing code. We acknowledge Rima Sch\"{u}ssler, Henrik Hirzler and Matteo Mazzanti for fruitful discussions.  This work was supported by the Netherlands Organization for Scientific Research (Grant Nos. 680.91.120 and 680.92.18.05, R.G.). A.S.N is supported by the Dutch Research Council (NWO/OCW), as part of the Quantum Software Consortium programme (project number 024.003.037).
\end{acknowledgments}

\bibliography{paper}

\end{document}